\documentclass[fleqn,twoside]{article}
\usepackage{amssymb}
\usepackage{latexsym}
\usepackage{espcrc2}
\usepackage{multicol}
\usepackage{axodraw}

\usepackage{graphicx}
\usepackage[figuresright]{rotating}
\usepackage{psfrag}
\usepackage{xspace}
\usepackage{epsfig}

\newcommand{\CnuB}{C\(\nu\)B\xspace}
\newcommand{\UHEnu}{UHE\(\nu\)\xspace}
\newcommand{\slache}[1]
{#1\!\!\!\!/}
\newcommand{\nubar}{{\bar{\nu}}}

\newcommand{\AmS}{{\protect\the\textfont2
  A\kern-.1667em\lower.5ex\hbox{M}\kern-.125emS}}

\title{Propagation of ultra-high energy neutrinos in the cosmic neutrino background}

\author{V. Van Elewyck\address[MCSD]{Institut de Physique Nucl\'eaire d'Orsay \\
15, rue G. Cl\'emenceau, 91406 Orsay Cedex, France \\
 Email: vero@ipno.in2p3.fr }%
        \thanks{supported by CNRS-IN2P3 and the European Community under Marie Curie Fellowship
        MEIF-CT-2005 025057. }
        }

\begin{document}

\begin{abstract}
UHE cosmic neutrino interaction with the cosmic neutrino background
(CnuB) is expected to produce absorption dips in the UHE neutrino
flux at energies above the threshold for Z-boson resonant
production. The observation of these dips would constitute an
evidence for the existence of the CnuB; they could also be used to
determine the value of the relic neutrino masses as well as some
features of the population of UHE neutrino sources. After breafly
discussing the current prospects for relic neutrino spectroscopy, we
present a calculation of the UHE neutrino transmission probability
based on finite-temperature field theory which takes into account
the thermal motion of the relic neutrinos. We then compare our
results with the approximate expressions existing in the literature
and discuss the influence of thermal effects on the absorption dips
in the context of realistic UHE neutrino fluxes and favoured
neutrino mass schemes.
\end{abstract}

\maketitle

\section{Introduction and motivations}

The existence of a cosmological background of relic neutrinos (\CnuB
), characterized by a present temperature of $T_{\nu 0} \approx
1.95$ K ($1.69 \times 10^{-4}$ eV) and a number density $n_{\nu 0}
\approx 56 \ \mathrm{cm}^{-3}$ per species, is a robust prediction
of Big Bang cosmology~\cite{peebleskolb}. Its presence, at least at
early times, seems now confirmed by data coming from observational
cosmology, which also provide bounds on the number of neutrino
species (now greater than zero) and on their absolute mass scale
(see \cite{hannestadpastor} for an extensive discussion of these
issues). Direct detection of the \CnuB at present time is however
much more problematic. Several laboratory experiments (discussed
i.e. in \cite{eberle}) have been proposed to detect relic neutrino
through their weak interactions, using torsion balances or even
accelerated beams of nuclei, but the technological improvements
required in those experimental setups are way beyond the state of
the art.

An interesting alternative is provided by exploiting cosmic rays -
actually ultra-high energy cosmic neutrinos (\UHEnu) - as a natural
beam and search for evidence of their interaction with the relic
neutrino background. Two approaches have emerged in that context:
the first one searches for "emission" features and the production of
charged cosmic rays or photons beyond the GZK cutoff, through the
so-called "Z-burst" mechanism \cite{fargion99,weiler99} (the
dominant \UHEnu - \CnuB interaction channel being $\nu \ \nubar
\longrightarrow Z \longrightarrow X$). This mechanism probes the
universe delimited by the GZK sphere and is therefore sensitive to
local overdensities in the \CnuB that may arise from the
gravitational clustering on massive structures, like the Virgo
cluster \cite{ringwald05}, and would result in a directional excess
of UHE cosmic rays.

The second approach, on which we will concentrate here, looks for
"absorption" features in the \UHEnu flux that would reflect their
interaction with the relic neutrinos along their path
\cite{weiler,roulet,yoshida,ringwald,quigg}. As illustrated below,
the position of the absorption dip approximately corresponds to the
redshifted resonance energy for the Z-boson exchange, namely
\begin{equation}
K_{res} \approx \frac{M_Z^2}{2m_\nu (1+z_s)}
\end{equation}
so that it directly depends on the value of the neutrino mass
$m_\nu$ and of the source redshift $z_s$. The shape and depth of the
dip also reflect to some extent the neutrino mixing pattern as well
as the characteristics and distribution of \UHEnu sources. In that
sense, and provided adequate sensitivity and energy resolution of
the detectors, a detailed spectroscopic study of the \UHEnu flux
could allow the determination of the absolute neutrino masses along
with an experimental proof of existence of the \CnuB
\cite{pas,fargion}. This mechanism has generated a renewed interest
in the light of the recent results obtained both in observational
cosmology and for determination of the neutrino mass and mixing
schemes.

However, most of the work in the literature describe the \UHEnu
-\CnuB interactions assuming that relic neutrinos are at rest, while
the effects of thermal motion should be included as soon as the
momentum of relic neutrinos becomes comparable to their mass (or
even before). Therefore we have chosen to address the question of
relic neutrino spectroscopy in the framework of finite-temperature
field theory (FTFT), which allows to take into account the thermal
effects in a systematic way \cite{ourpaper}. In section
\ref{sec:damping}, we briefly present our calculation of the damping
of an UHE neutrino travelling across the \CnuB and determine the
absorption probability for a neutrino emitted at a given redshift.
 We then illustrate our
calculations in two realistic contexts and explore various
combinations of parameters to investigate the differences between
the FTFT calculation and previous approximations. In section
\ref{sec:flux} we discuss how the thermal broadening of the
absorption lines in the \UHEnu flux could affect the determination
of $m_\nu$ and of the characteristics of the \UHEnu source
population. In section \ref{sec:cluster} we comment our results in
the context of relic neutrino clustering, for different hypothesis
on the density and scale of the clusters. Concluding remarks are
presented in section \ref{sec:conclusion}.

\section{Damping  of UHE neutrinos across the relic neutrino background}
\label{sec:damping}
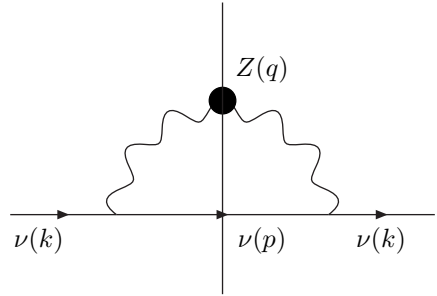
\begin{figure}[bt!]
\begin{center}
\begin{picture}(170,130)(-85,-30)
\ArrowLine(40,0)(80,0) \Text(60,-10)[c]{$\nu(k)$}
\ArrowLine(-40,0)(40,0) \Text(15,-10)[c]{$\nu(p)$}
\ArrowLine(-80,0)(-40,0) \Text(-60,-10)[cr]{$\nu(k)$}
\PhotonArc(0,0)(40,0,180){4}{6.5} \GCirc(0,43){5}{0}
\Text(15,50)[cb]{$Z(q)$} \Line(0,-30)(0,80)
\end{picture}
\caption{\footnotesize{Feynman diagram for the one-loop self-energy of an UHE neutrino due to a 
Z-boson exchange with an (anti-)neutrino from the relic background; the blob on the Z propagator 
indicates that we use the dressed propagator and the cut is to select the imaginary part of the 
diagram.}} \label{fig:feynman}
\end{center}
\end{figure}

 The equation of motion of an UHE neutrino with
four-momentum $k^\mu = (\mathcal{E}_{_K},\vec{K})$ and mass $m_\nu$
traveling across the \CnuB reads
\begin{equation}
\label{eq:motion}
 (\slache{k} - m_\nu - \Sigma)\psi = 0
\end{equation}
 where the
self-energy $\Sigma$ accounts for the interactions with the
surrounding medium. The dominant process is here the
 Z-boson exchange in the s-channel, as shown in fig. \ref{fig:feynman}. We determine $\Sigma$ from a FTFT
 one-loop calculation carried out in terms of the (vacuum) Z 
propagator and the thermal propagator of the relic neutrinos. 
The latter depends on the functions $f_\nu(P)$ and $f_{\bar{\nu}}(P)$ which describe the 
momentum distributions of neutrinos (antineutrinos) in the thermal bath. These functions take the 
simple relativistic Fermi-Dirac form \begin{equation} f_\nu(P) =
f_{\bar{\nu}}(P)=\frac{1}{(e^{P/T_\nu}+1)},
\end{equation} where $T_{\nu}$
is the temperature of the \CnuB and we have neglected the chemical
potential. It is worth noting here that, although relic neutrinos
are not relativistic anymore at present time, their distribution has
maintained the form it had at the time of neutrino decoupling,
corresponding to $T_\nu \sim 1$ MeV.

\begin{figure*}
\label{fig:}
 \psfrag{20}[c]{\tiny \phantom{n} {$_2$}}
 \psfrag{40}[c]{\tiny \phantom{n} {$_4$}}
 \psfrag{60}[c]{\tiny \phantom{n}{$_6$}}
 \psfrag{80}[c]{\tiny \phantom{n}{$_8$}}
\psfrag{200}[c]{\tiny \phantom{n}{$_2$}}
 \psfrag{400}[c]{\tiny \phantom{n}{$_4$}}
 \psfrag{600}[c]{\tiny \phantom{n}{$_6$}}
 \psfrag{800}[c]{\tiny \phantom{n}{$_8$}}
\psfrag{1000}[c]{\tiny \phantom{n}{}} \psfrag{2000}[c]{\tiny
\phantom{n}{$_2$}}
 \psfrag{4000}[c]{\tiny \phantom{n}{$_4$}}
 \psfrag{6000}[c]{\tiny \phantom{n}{$_6$}}
\psfrag{100}[c]{\tiny \phantom{n}{}} \psfrag{2}[c]{\tiny
\phantom{nn} \raisebox{0.1cm}{$_2$}} \psfrag{4}[c]{\tiny
\phantom{nn} \raisebox{0.1cm}{$_4$}} \psfrag{6}[c]{\tiny
\phantom{nn} \raisebox{0.1cm}{$_6$}}
 \psfrag{1.5}[c]{\tiny \phantom{nnn} \raisebox{0.1cm}{$_{1.5}$}}
 \psfrag{1}[c]{\tiny \phantom{nn} \raisebox{0cm}{$_1$}}
 \psfrag{0.5}[c]{\tiny \phantom{nnn} \raisebox{0cm}{$_{0.5}$}}
 \psfrag{0}[c]{\tiny \raisebox{0.1cm}{$_0$}}
 \psfrag{A}[c]{\tiny \phantom{mmmmmmmmmm}\raisebox{-0.2cm}{$P\,\, [10^{-3}\,\mathrm{eV}]$}}
\psfrag{BBB}[c]{\tiny \phantom{mmmmm}\raisebox{0cm}{$K\,\,
[10^{24}\,\mathrm{eV}]$}} \psfrag{BB}[c]{\tiny
\phantom{mmmmm}\raisebox{-0.2cm}{$K\,\, [10^{23}\,\mathrm{eV}]$}}
\psfrag{B}[c]{\tiny \phantom{mmmmm}\raisebox{-0.2cm}{$K\,\,
[10^{22}\,\mathrm{eV}]$}}
 \psfrag{27}[c]{\tiny \phantom{n} \raisebox{0cm}{$10^{27}$}}
 \psfrag{26}[c]{\tiny \phantom{n} \raisebox{0cm}{$10^{26}$}}
 \psfrag{25}[c]{\tiny \phantom{n} \raisebox{0cm}{$10^{25}$}}
 \psfrag{24}[c]{\tiny \phantom{n} \raisebox{0cm}{$10^{24}$}}
 \psfrag{23}[c]{\tiny \phantom{n} \raisebox{0cm}{$10^{23}$}}
 \psfrag{22}[c]{\tiny \phantom{n} \raisebox{0cm}{$10^{22}$}}
 \psfrag{21}[c]{\tiny \phantom{n} \raisebox{0cm}{$10^{21}$}}
 \psfrag{1.}[c]{\tiny \phantom{} \raisebox{0.1cm}{$1$}}
 \psfrag{0.8}[c]{\tiny \phantom{} \raisebox{0.1cm}{$0.8$}}
 \psfrag{0.6}[c]{\tiny \phantom{} \raisebox{0.1cm}{$0.6$}}
 \psfrag{0.4}[c]{\tiny \phantom{} \raisebox{0.1cm}{$0.4$}}
 \psfrag{0.2}[c]{\tiny \phantom{} \raisebox{0.1cm}{$0.2$}}
 \psfrag{0.}[c]{\tiny \phantom{} \raisebox{0.1cm}{$0$}}
 \psfrag{X}[c]{\tiny \raisebox{-0.5cm}{$K_0 [\mathrm{eV}]$}}
 \psfrag{Y}[c]{\tiny {$P_\mathrm{T}$}}
 \psfig{file=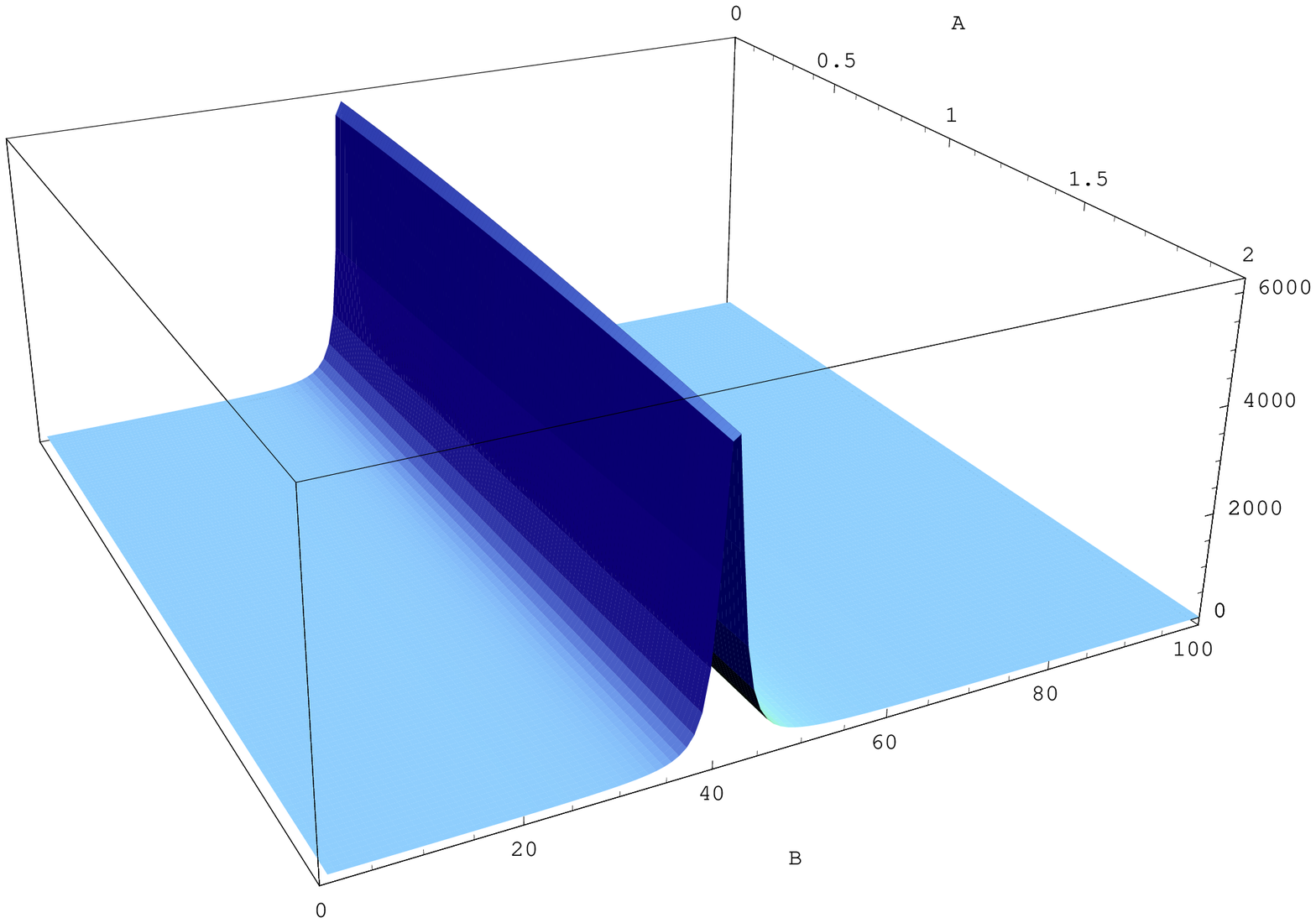,width=.27\textwidth}
 \hfill
 \psfig{file=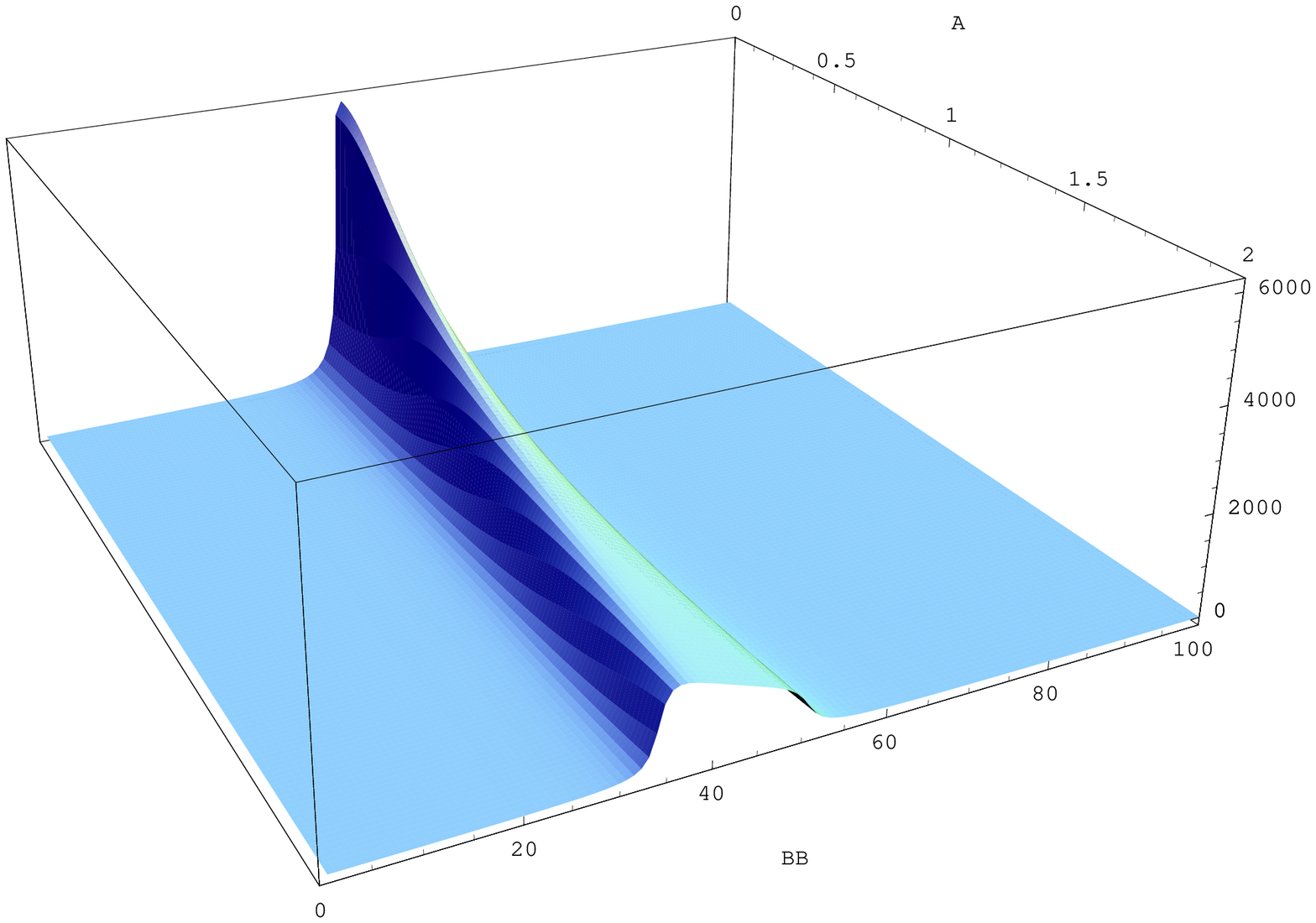,width=.27\textwidth,angle=0}
 \hfill
 \psfig{file=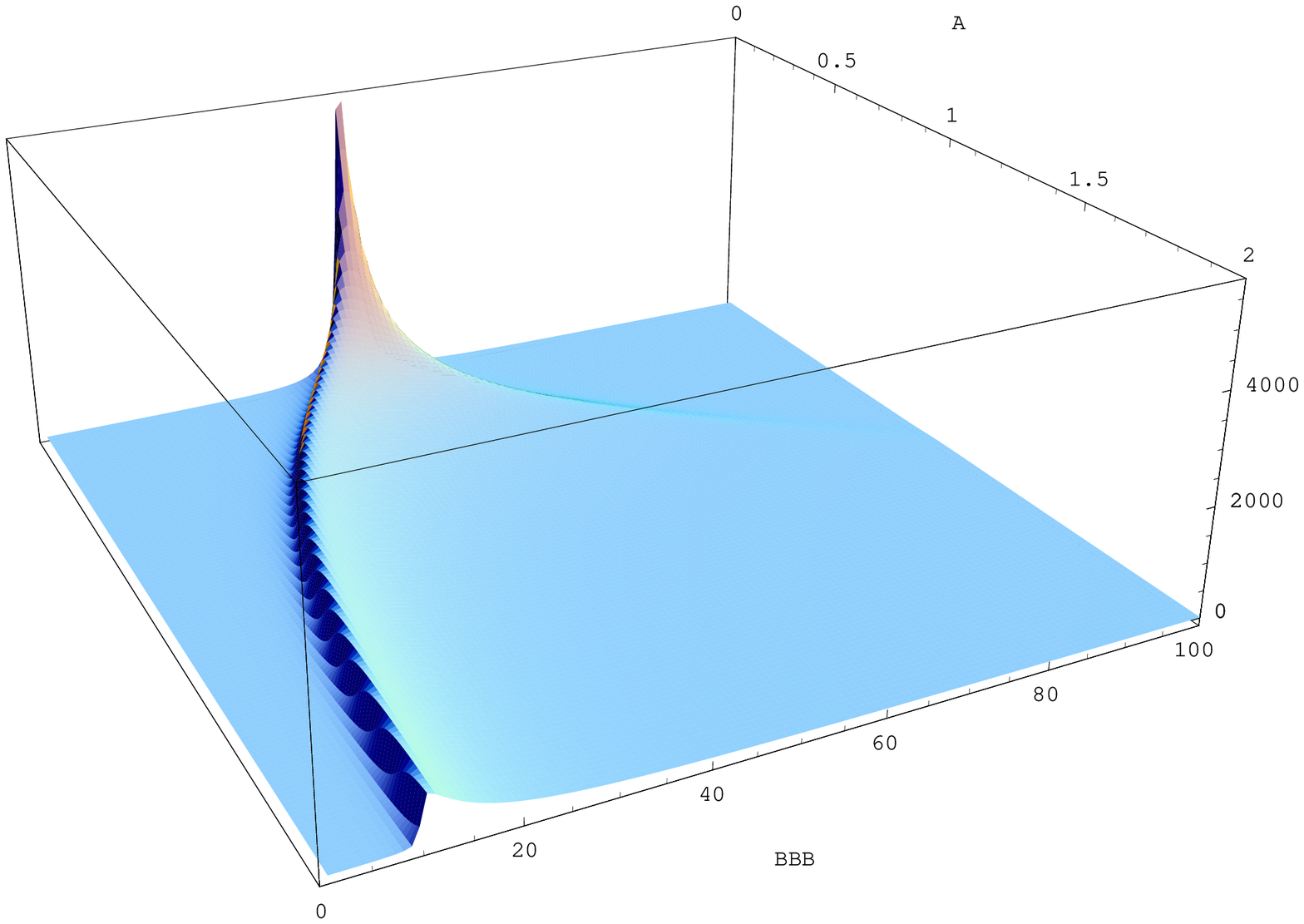,width=.27\textwidth,angle=0} \\[3ex]
 \psfig{file=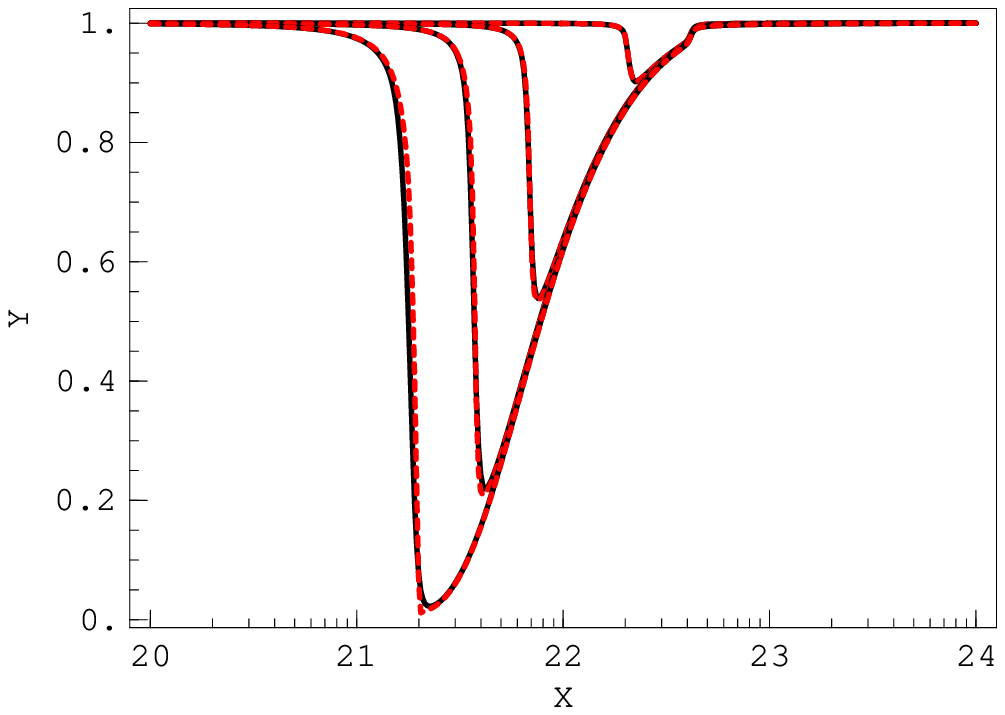,width=0.22\textwidth,angle=0}
 \hfill
 \psfig{file=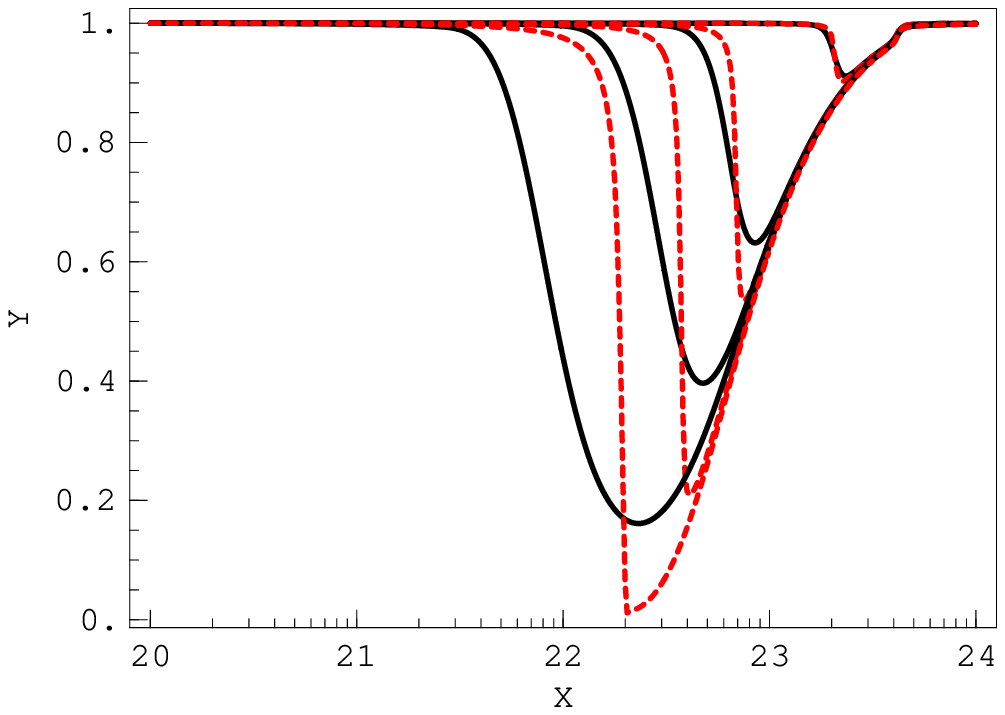,width=0.22\textwidth,angle=0}
 \hfill
 \psfig{file=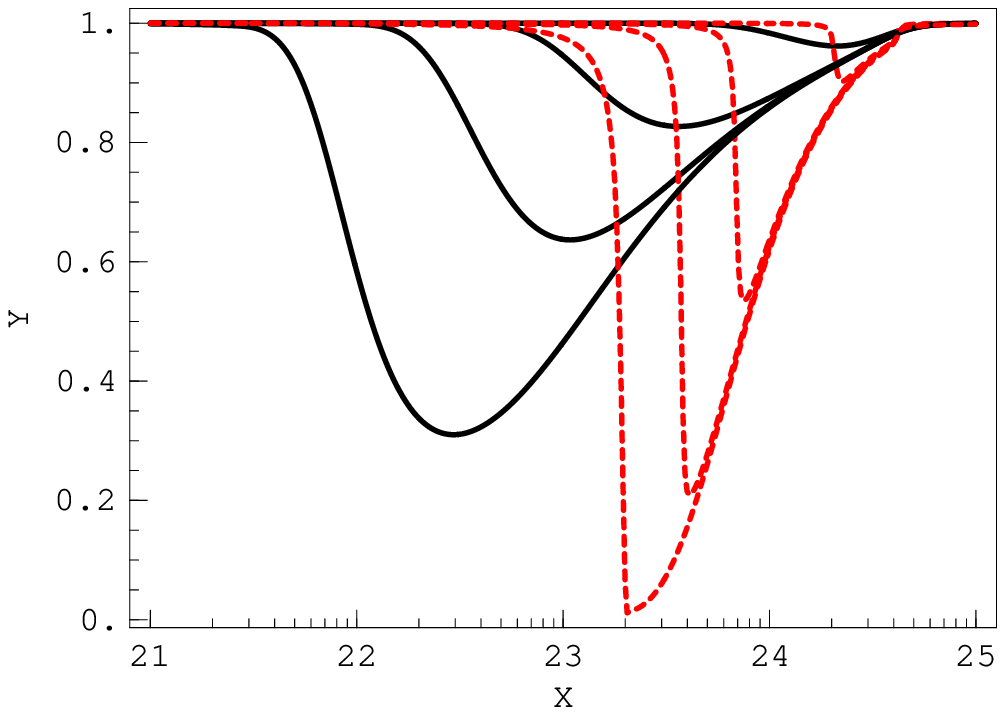,width=0.22\textwidth,angle=0}
 \caption{ \footnotesize{ {\bf Top :} cross-section $\sigma_{\nu\bar{\nu}}(P,K)$, in units of 
$10^{-31} cm^2$, as given by eq.~(6), as a function of the energy of the incident neutrino, $K$, 
and of the relic neutrino momentum, $P$. From left to right, the panels correspond to a neutrino 
mass $10^{-1}$, $10^{-2}$, and $10^{-3}$ eV.
{\bf Bottom :} Transmission probability $P_\mathrm{T}(K_0,z_\mathrm{s})$ as a function of the 
\UHEnu energy as detected on Earth, $K_0$, for a source located at redshifts $z_\mathrm{s}=1$, 
$5$, $10$, $20$ (from top to bottom in each panel) and for a neutrino mass $m_\nu = 10^{-1}$, 
$10^{-2}$, $10^{-3}$ eV (from left to right). The continued, black curves correspond to the full 
damping as given by eqs.~(5) and (6), while the dotted (red) curves are for the approximation 
of relic neutrinos at rest, eq.(7). } }
 \end{figure*}

The dispersion relation corresponding to eq. (\ref{eq:motion}) is 
given by $\mathcal{E}_K = \mathcal{E} _r(K) - i\,\gamma (K)/2$,
where $\mathcal{E}_K$ and $\gamma$ are functions of K.
The damping factor $\gamma$ governs the propagation of the \UHEnu across the background of relic 
neutrinos and is directly related to the imaginary part of the self-energy, $\Sigma_{\,i}$ 
\cite{dolivo}. In the approximation that the \UHEnu are ultrarelativistic and that we can neglect the 
background effects on their energy ($\mathcal{E} _r (K)\simeq K$), the damping can be written as 
(see \cite{ourpaper} for the detailed calculation)
\begin{eqnarray}
\label{eq:gammaUR} \gamma_{\nu\bar{\nu}}(K) &=& 
-\frac{1}{K}\left.\mathop{\mathrm{Tr}}(\slache{k}\Sigma_i)\right|_{\mathcal{E}_r=K}\\
&=& \int_0^\infty \frac{dP}{2\pi^2} \ P^2 \ f_{\bar{\nu}} (P) \
\sigma_{\nu\bar{\nu}} (P,K). \label{eq:gamma}
\end{eqnarray}
 For $m_\nu \ll M_Z,K$ and neglecting terms of order $\Gamma_Z^2/M_Z^2$, we have
\begin{eqnarray}
\sigma_{\nu\bar{\nu}}(P,K) &=& \frac{2\sqrt{2}G_\mathrm{F}\Gamma_Z
M_Z}{2KE_p} \left\{ 1 + \frac{M_Z^2}{4KP} \right. \nonumber\\
&& \hspace{-2.4cm} \times \ln\left(\frac{4K^2(E_p + P)^2 -
4M_Z^2K(E_p+P)+M_Z^4}{4K^2(E_p - P)^2 -
4M_Z^2K(E_p-P)+M_Z^4}\right) \nonumber \\
&& \nonumber\\
&& \hspace{-2.4cm} + \frac{M_Z^3}{4KP
\Gamma_Z}\left[\arctan\left(\frac{2K(E_p+P)-M_Z^2}{\Gamma
M_Z}\right) \right. \nonumber\\
&& \hspace{-1.6cm}\left.\left.
-\arctan\left(\frac{2K(E_p-P)-M_Z^2}{\Gamma
M_Z}\right)\right]\right\}. \label{eq:sigma}
\end{eqnarray}
where $E_p = \sqrt{P^2 + m_\nu^2}$ is the energy of the relic neutrino. Taking the limit of 
eq.~(\ref{eq:sigma}) for $P \rightarrow 0$, one recovers the approximated cross-section used for 
relic neutrinos at rest, with the Z peak at the \UHEnu "bare" resonance energy $K_{res} = 
M_Z^2/(2m_\nu)$. The damping reads in that case
\begin{eqnarray}
\gamma_{\nu\bar{\nu}}^0(K)  &=& 2\sqrt{2} G_\mathrm{F} \Gamma_Z M_Z\ n_\nu \ \nonumber\\
&& \hspace{-0.3cm}\times \frac{2 K m}{4K^2 m^2 - 4M_Z^2K m +M_Z^4} \ , \label{eq:gammaapprox}
\end{eqnarray}
in agreement with the results of \cite{roulet}. The expressions used
in \cite{ringwald} can be obtained by further evaluating the
cross-section at the pole of the resonance, $2mK_{res} = M_Z^2$
(narrow-width approximation).
 However, this approximation is no longer valid for relatively small $m_\nu$ and/or large \CnuB
 temperatures: fig.~1 (top line) 
shows how the resonance peak in the $\nu \bar{\nu} \rightarrow Z$ cross-section broadens and 
shifts to lower \UHEnu energies as $P$ increases. Actually two
effects combine: the modification of the cross-section peak due to
its dependance in $E_p$, and the thermal distribution which selects
a range of relic neutrino momenta close to the temperature of the
\CnuB. The net effect of this thermal broadening is a reduction of
the damping, which affects the transmission probability and the
depth and shape of the absorption dips.

The transmission probability for an \UHEnu 
emitted at a redshift $z_s$ to be detected on Earth with an energy $K_0$ is obtained by 
integrating the damping along the \UHEnu path, taking into account that both the \UHEnu energy and 
the \CnuB temperature are redshifted:
\begin{eqnarray}
P_\mathrm{T}(K_0,z_\mathrm{s})&=& \nonumber\\
&& \hspace{-2.1cm}\exp\left[{-\int_0^{z_\mathrm{s}} \frac{dz}{H(z)(1+z)} 
\gamma_{\nu\bar{\nu}}(K_0(1+z))}\right], \label{eq:PTredshift}
\end{eqnarray}
where $H = H_0\ \sqrt{0.3(1+z)^3 +0.7}$ is the Hubble factor as
suggested by recent observations \cite{spergel03}. Fig.~1 (bottom
line) compares the transmission probabilities obtained from eqs.
(\ref{eq:gamma}) and (\ref{eq:gammaapprox}), for an \UHEnu emitted
at different redshifts and $m_\nu$ ranging from $10^{-1}$ to
$10^{-3}$ eV. As long as $m_\nu / T_\nu \geq 10^{-2}$, the shape of
the absorption dip is not affected by thermal broadening and is
rather sharply delimited, at high energies, by the bare resonant
energy for the propagating neutrino, $K_0=K_{res} = M_Z^2/(2m_\nu)$,
and at low energies by the redshifted resonant energy
$K_0=K_{res}/(1+z_s)$. Evaluating the position of these points would
in principle allow us to determine $m_\nu$ as well as $z_s$, the
redshift at which the \UHEnu was emitted. As $m_\nu / T_\nu$
decreases, however, the absorption dips get broadened and shift to
lower energies, and this effect also increases with the redshift
since UHE neutrinos from distant sources are emitted in a hotter
background.

\section{Absorption lines in the UHE neutrino flux}
\label{sec:flux} The results presented so far deal with a
monoenergetic source of \UHEnu located at a given redshift. A
realistic situation would more probably involve a distribution of
sources which emit \UHEnu with some given energy spectrum and flavor
composition, the latter evolving along the neutrino pathway. To
investigate these effects, and following the approach of
\cite{ringwald}, we have considered a flux of \UHEnu of the form:
\begin{eqnarray}
\label{eq:flux}
\mathcal{F}_\nu(K_0) &\hspace{-0.2cm}=& \hspace{-0.2cm}\frac{1}{4 \pi} \int_0^\infty \frac{dz}{H(z)} \ P_\mathrm{T}(K_0,z) \ 
\eta(z)\ J_\nu(K_0).so \nonumber
\end{eqnarray}
The distribution of sources,
\begin{equation}
\eta(z)=\eta_{\,0} \,(1+z)^n\, \theta(z-z_\mathrm{min})\, 
\theta(z_\mathrm{max}-z), \label{eq:eta}
\end{equation}
is well-suited for an approximate description of models ranging from
astrophysical production sites ("bottom-up" mechanisms for which $n
\simeq 4$ and $z_{max} \leq 10$) to exotic, non-accelerator sources
(which typically have $n \simeq 1-2$ and may extend to larger
redshifts). The injection spectrum is taken as a power-law with a
cutoff at some high energy $K_{max} > K_{res}(1+z)$ (we do not
consider here the possibility of broken power-law spectra):
\begin{equation}
J_\nu(K)= j_\nu \, K^{-\alpha}\, 
\theta(K_\mathrm{max}-K) \label{eq:J}
\end{equation}
 with the
spectral index $\alpha$ ranging between 1 and 2, depending on the
production mechanism considered. Under these assumptions, the
normalized flux only depends on the difference $n - \alpha$ of the
spectral indexes. We then computed numerically the
 \UHEnu flux produced for two distinct values of the spectral index combination:
$n-\alpha = 2$, which is representative of a distribution of
astrophysical sources, and $n - \alpha =0$, which could describe the
\UHEnu flux produced through some top-down process.

As for the flavor content of $\mathcal{F}_\nu$, the ratio predicted
for standard, hadronic sources is $J_{\nu_e} : J_{\nu_\mu} :
J_{\nu_\tau} = 1:2:0$, while more exotic mechanisms could be
characterized by a democratic $J_{\nu_e} : J_{\nu_\mu} :
J_{\nu_\tau} = 1:1:1$. In both cases, and in view of our current
knowledge of the neutrino mixing parameters, it is reasonable to
think that the total neutrino flux detected at Earth will just be
the sum on all three neutrino mass eigentstates \cite {ringwald}.
Therefore we considered mass patterns compatibles with the favoured
3-neutrino hierarchical schemes \cite{bell}, namely:
\[\begin{array}{ll}
normal: & m_3 = 5 \ 10^{-2}\ eV,\ m_2 = 9 \ 10^{-3}\ eV \\
inverted: & m_3 = m_2 = 5 \ 10^{-2}\ eV
\end{array}\]
and chose the values $m_1 = 10^{-3}$ eV or $10^{-4}$ eV for the
third, unknown mass.

\begin{figure*}
\label{fig:fluxTD}
 \psfrag{27}[c]{\tiny \phantom{n} \raisebox{0cm}{$10^{27}$}}
 \psfrag{26}[c]{\tiny \phantom{n} \raisebox{0cm}{$10^{26}$}}
 \psfrag{25}[c]{\tiny \phantom{n} \raisebox{0cm}{$10^{25}$}}
 \psfrag{24}[c]{\tiny \phantom{n} \raisebox{0cm}{$10^{24}$}}
 \psfrag{23}[c]{\tiny \phantom{n} \raisebox{0cm}{$10^{23}$}}
 \psfrag{22}[c]{\tiny \phantom{n} \raisebox{0cm}{$10^{22}$}}
 \psfrag{21}[c]{\tiny \phantom{n} \raisebox{0cm}{$10^{21}$}}
 \psfrag{20}[c]{\tiny \phantom{n} \raisebox{0cm}{$10^{20}$}}
 \psfrag{1.}[c]{\tiny \phantom{} \raisebox{0.1cm}{$1$}}
 \psfrag{0.9}[c]{\tiny \phantom{} \raisebox{0.1cm}{$0.9$}}
 \psfrag{0.8}[c]{\tiny \phantom{} \raisebox{0.1cm}{$0.8$}}
 \psfrag{0.7}[c]{\tiny \phantom{} \raisebox{0.1cm}{$0.7$}}
 \psfrag{0.6}[c]{\tiny \phantom{} \raisebox{0.1cm}{$0.6$}}
 \psfrag{0.4}[c]{\tiny \phantom{} \raisebox{0.1cm}{$0.4$}}
 \psfrag{0.2}[c]{\tiny \phantom{} \raisebox{0.1cm}{$0.2$}}
 \psfrag{0.}[c]{\tiny \phantom{} \raisebox{0.1cm}{$0$}}
 \psfrag{X}[c]{\tiny \raisebox{-0.5cm}{$K_0 [\mathrm{eV}]$}}
 \psfrag{Y}[c]{\tiny {}}
 \psfig{file=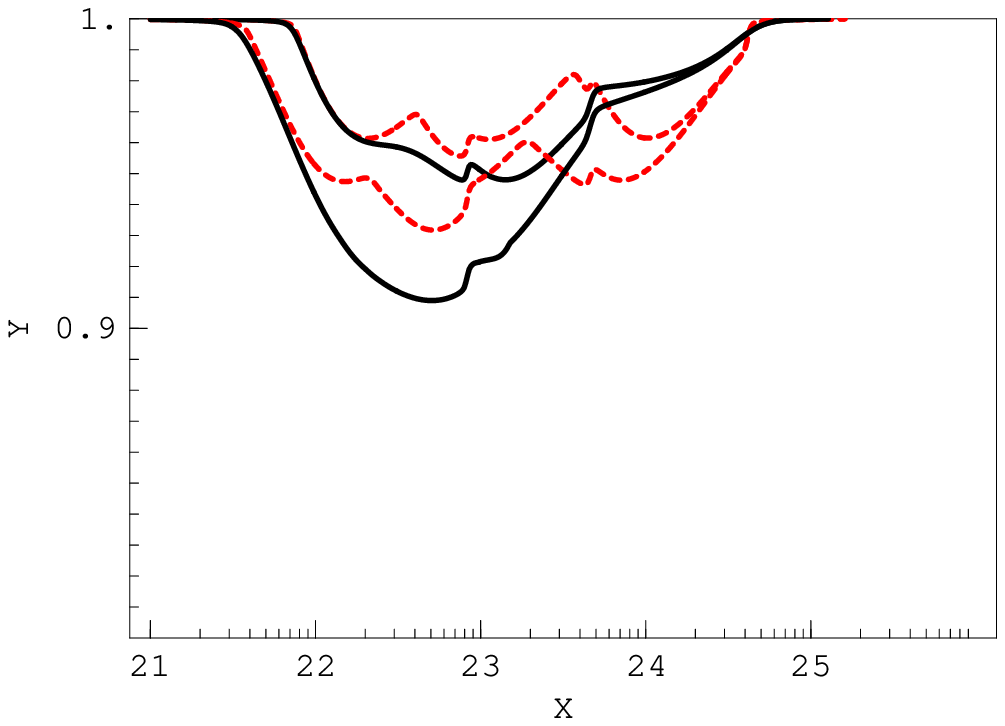,width=0.30\textwidth,angle=0}
 \hspace{4cm}
 \psfig{file=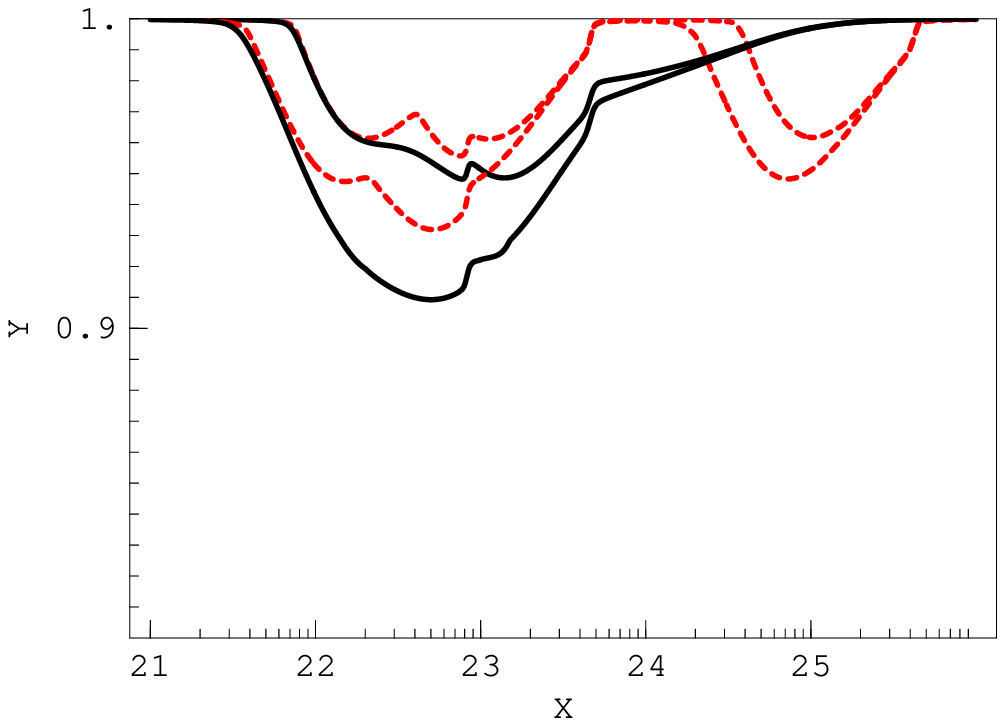,width=0.30\textwidth,angle=0}
 \hfill\\[3ex]
\hfill
\psfig{file=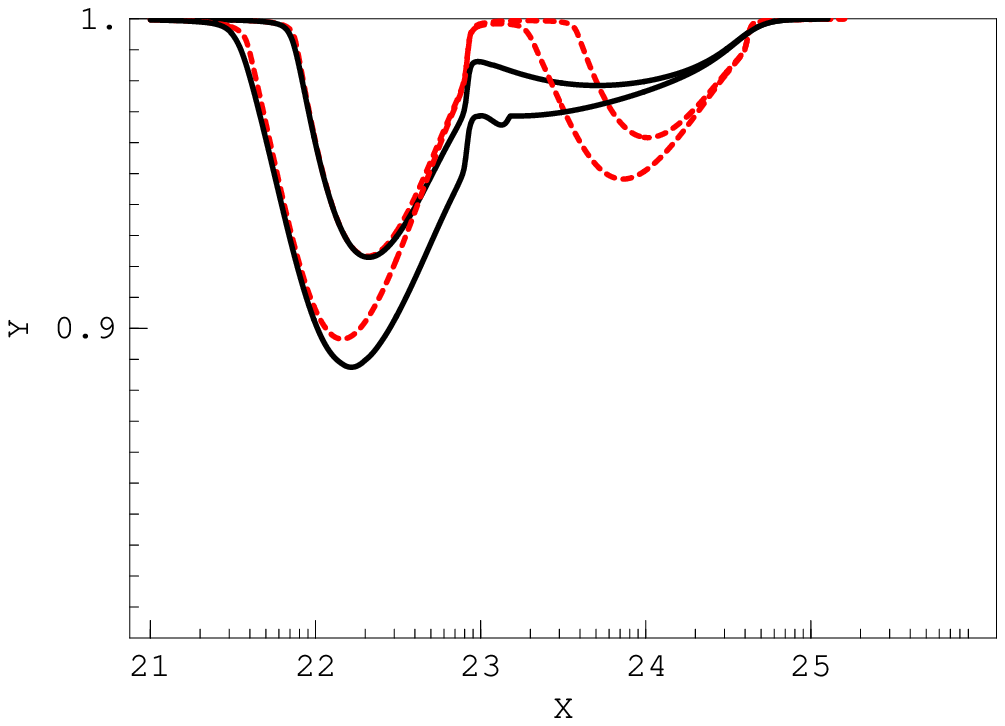,width=0.30\textwidth,angle=0}
\hspace{4cm}
\psfig{file=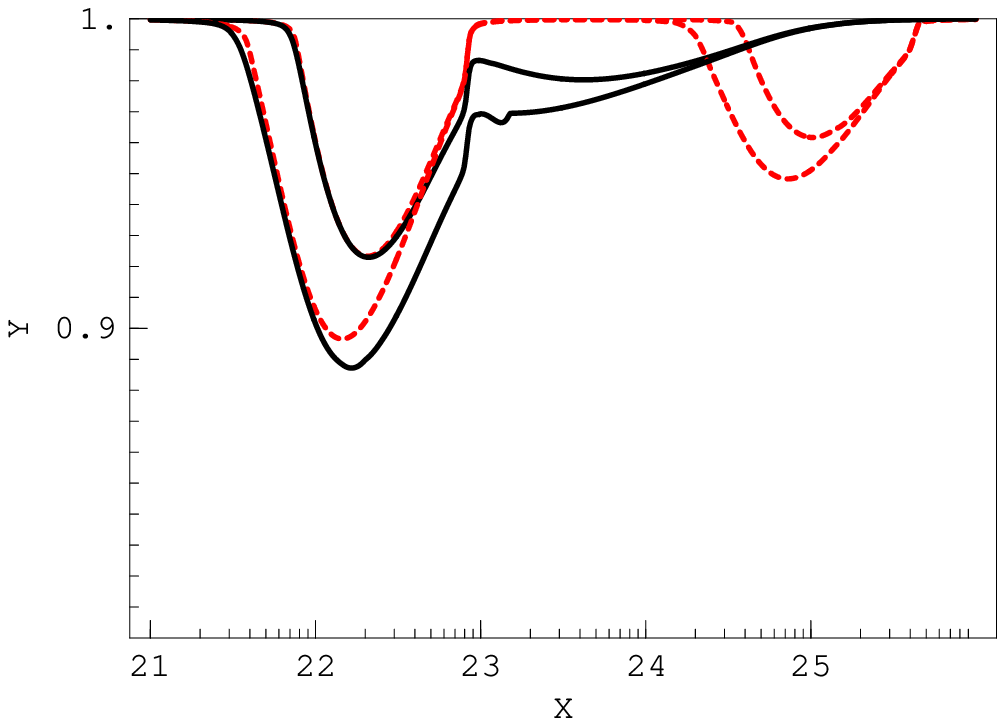,width=0.30\textwidth,angle=0}
 \caption{ \footnotesize{ \UHEnu flux in presence of damping, $\mathcal{F}_\nu$, with $\alpha-n = 0$ and $z_s = 
10,20$ (typical top-down model), normalized to the 
corresponding flux in absence of interactions. The top row is for the normal neutrino mass hierarchy
and the bottom one for the inverted hierarchy. Plots on the left have $m_1 = 10^{-3}$ eV and plots
on the right $m_1 = 10^{-4}$ eV. Colour code 
is as in fig. 1.}}
 \end{figure*}

The results  are presented in figs. \ref{fig:fluxTD} and 4, which
display the all-flavour \UHEnu flux as a function of the present
energy $K_0$ of the UHE neutrino, normalized to the flux in absence
of absorption effects. Both figures show how thermal broadening
globally modify the shape and extension of the absorption lines
respect to the approximated case. The dip corresponding to the
smallest mass is almost always washed out and only contribute to
further broadening the line at high energies. For very large
$z_{max}$, the merging of the two other dips in the normal hierarchy
scheme makes it more difficult to differentiate from the inverted
one.
 \begin{figure*}
\label{fig:fluxBU}
 \psfrag{27}[c]{\tiny \phantom{n} \raisebox{0cm}{$10^{27}$}}
 \psfrag{26}[c]{\tiny \phantom{n} \raisebox{0cm}{$10^{26}$}}
 \psfrag{25}[c]{\tiny \phantom{n} \raisebox{0cm}{$10^{25}$}}
 \psfrag{24}[c]{\tiny \phantom{n} \raisebox{0cm}{$10^{24}$}}
 \psfrag{23}[c]{\tiny \phantom{n} \raisebox{0cm}{$10^{23}$}}
 \psfrag{22}[c]{\tiny \phantom{n} \raisebox{0cm}{$10^{22}$}}
 \psfrag{21}[c]{\tiny \phantom{n} \raisebox{0cm}{$10^{21}$}}
 \psfrag{20}[c]{\tiny \phantom{n} \raisebox{0cm}{$10^{20}$}}
 \psfrag{1.}[c]{\tiny \phantom{} \raisebox{0.1cm}{$1$}}
 \psfrag{0.9}[c]{\tiny \phantom{} \raisebox{0.1cm}{$0.9$}}
 \psfrag{0.8}[c]{\tiny \phantom{} \raisebox{0.1cm}{$0.8$}}
 \psfrag{0.7}[c]{\tiny \phantom{} \raisebox{0.1cm}{$0.7$}}
 \psfrag{0.6}[c]{\tiny \phantom{} \raisebox{0.1cm}{$0.6$}}
 \psfrag{0.4}[c]{\tiny \phantom{} \raisebox{0.1cm}{$0.4$}}
 \psfrag{0.2}[c]{\tiny \phantom{} \raisebox{0.1cm}{$0.2$}}
 \psfrag{0.}[c]{\tiny \phantom{} \raisebox{0.1cm}{$0$}}
 \psfrag{X}[c]{\tiny \raisebox{-0.5cm}{$K_0 [\mathrm{eV}]$}}
 \psfrag{Y}[c]{\tiny {}}
 \psfig{file=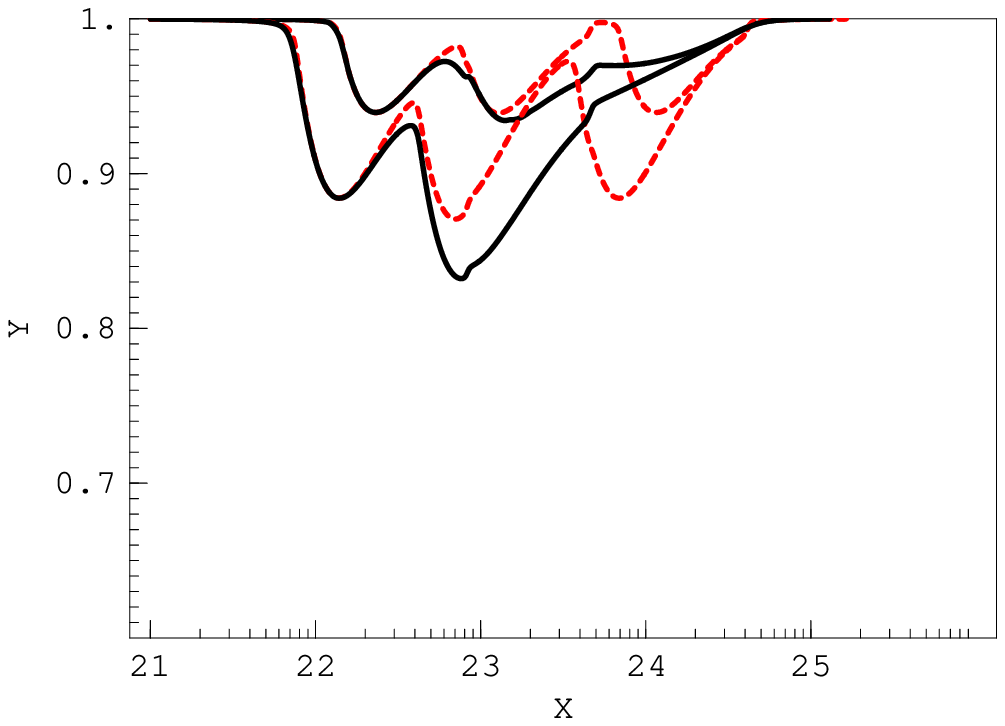,width=0.30\textwidth,angle=0}
 \hspace{4cm}
 \psfig{file=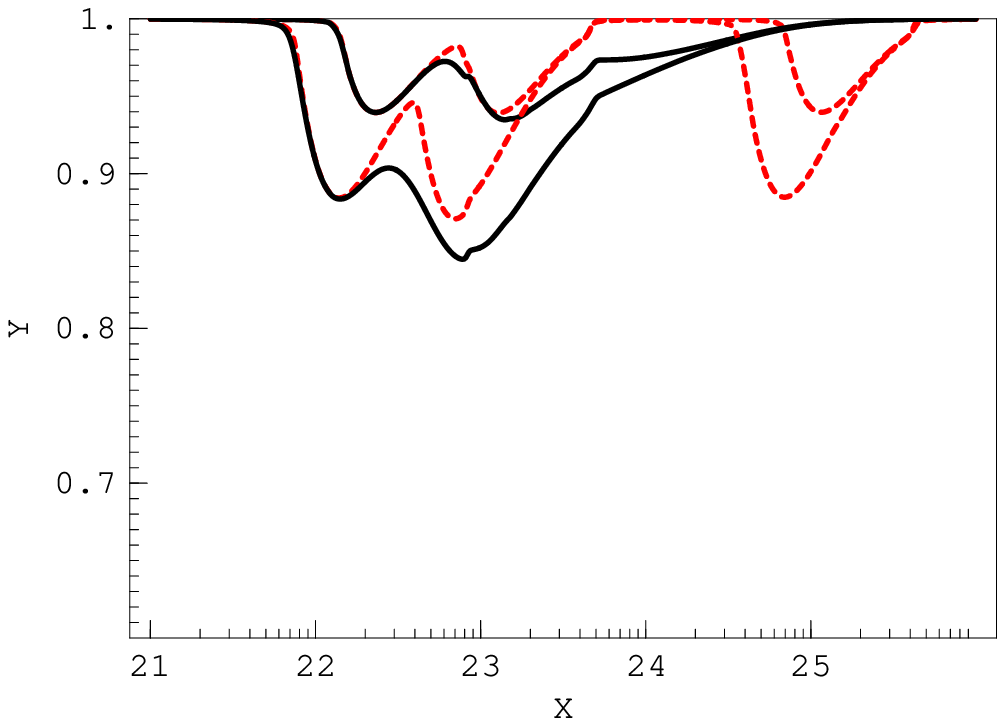,width=0.30\textwidth,angle=0}
 \hfill \\[3ex]
 \hfill
 \psfig{file=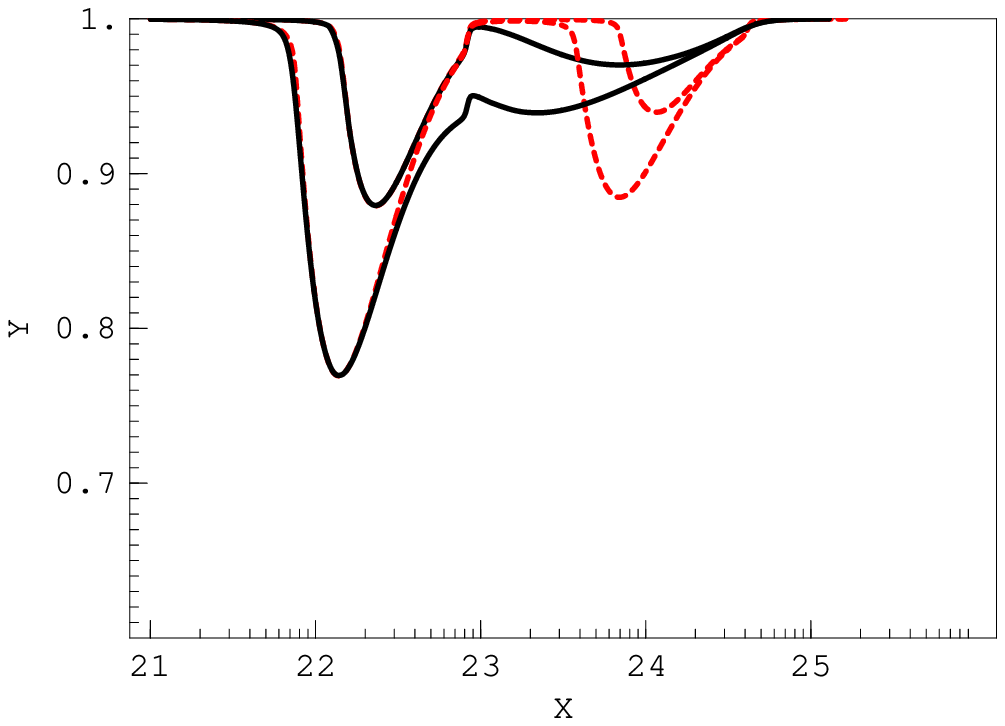,width=0.30\textwidth,angle=0}
 \hspace{4cm}
 \psfig{file=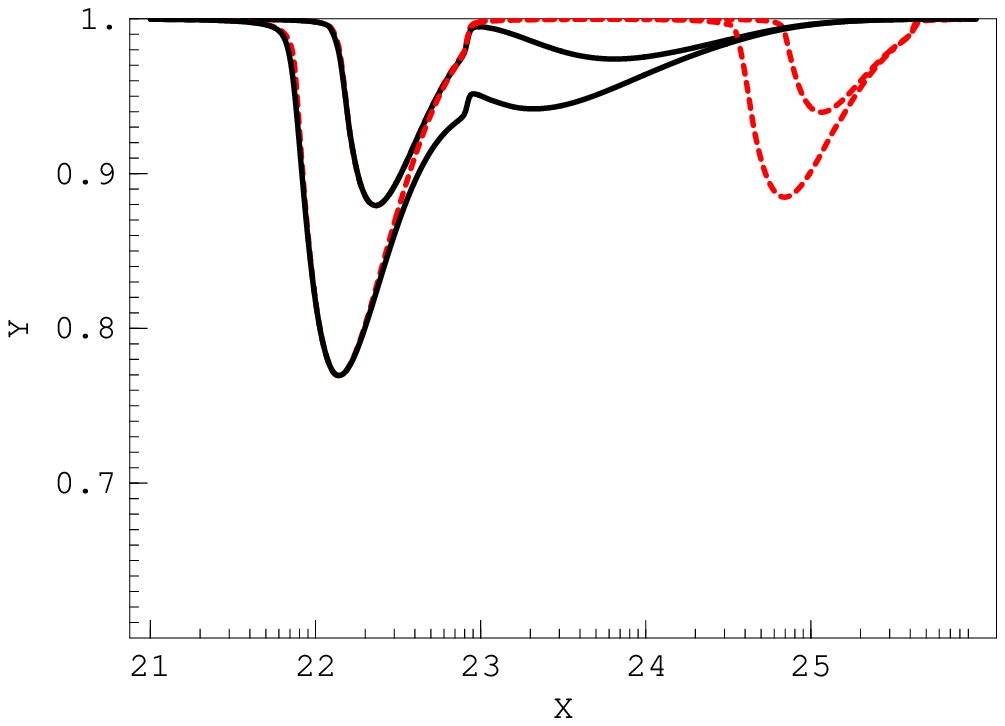,width=0.30\textwidth,angle=0}
 \caption{ \footnotesize{
\UHEnu flux in presence of damping, $\mathcal{F}_\nu$, with $\alpha-n = 2$ and $z_s = 
5, 10$ (typical bottom-up model), normalized to the 
corresponding flux in absence of interactions. The top row is for
the normal neutrino mass hierarchy and the bottom one for the
inverted hierarchy. Plots on the left have $m_1 = 10^{-3}$ eV and
plots
on the right $m_1 = 10^{-4}$ eV. Colour code 
is as in fig. 1.}}
 \end{figure*}
 \label{sec:cluster}

\begin{figure*}
 \begin{center}
 \psfrag{23}[c]{\tiny \phantom{n} \raisebox{0cm}{$10^{23}$}}
 \psfrag{22}[c]{\tiny \phantom{n} \raisebox{0cm}{$10^{22}$}}
 \psfrag{21}[c]{\tiny \phantom{n} \raisebox{0cm}{$10^{21}$}}
 \psfrag{20}[c]{\tiny \phantom{n} \raisebox{0cm}{$10^{20}$}}
 \psfrag{1.}[c]{\tiny \phantom{} \raisebox{0.1cm}{$1$}}
 \psfrag{0.8}[c]{\tiny \phantom{} \raisebox{0.1cm}{$0.8$}}
 \psfrag{0.6}[c]{\tiny \phantom{} \raisebox{0.1cm}{$0.6$}}
 \psfrag{0.4}[c]{\tiny \phantom{} \raisebox{0.1cm}{$0.4$}}
 \psfrag{0.2}[c]{\tiny \phantom{} \raisebox{0.1cm}{$0.2$}}
 \psfrag{0.}[c]{\tiny \phantom{} \raisebox{0.1cm}{$0$}}
 \psfrag{X}[c]{\tiny \raisebox{-0.5cm}{$K_0 [\mathrm{eV}]$}}
 \psfrag{Y}[c]{\tiny {$P_\mathrm{T}$}}
 \psfig{figure=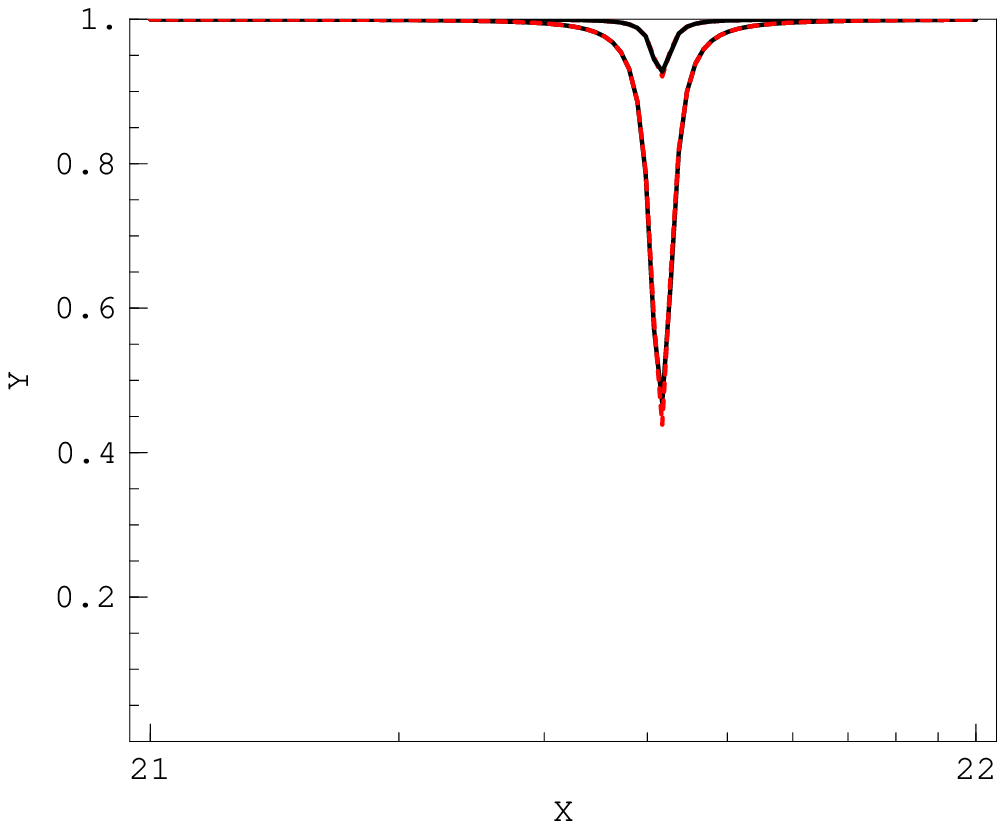,width=0.32\textwidth,angle=0}
 \hspace{2cm}
 \psfig{figure=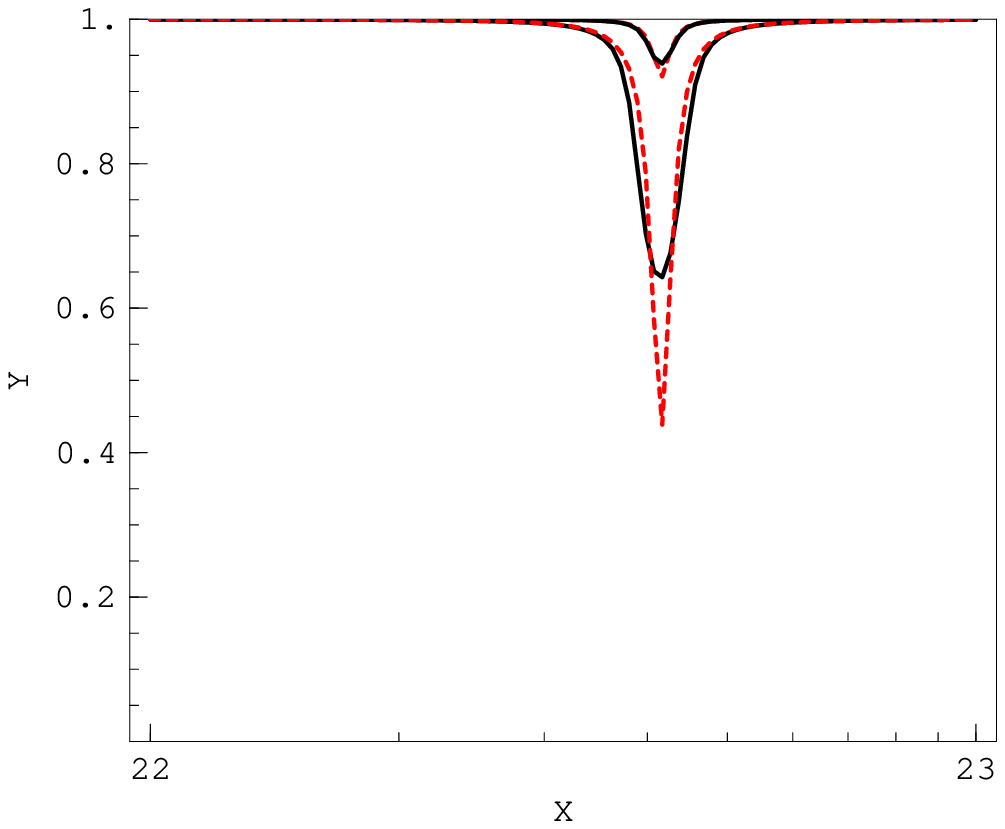,width=0.32\textwidth,angle=0}
 \hfill
 \caption{\footnotesize{Transmission probability for a cluster of extension 1 Mpc, made of 
neutrinos of mass 1 eV (left) and $10^{-1}$ eV (right), with a constant neutrino density 
$n^\mathrm{cl}_{\nu} = 10^3\ n_{0\nu}$ and $n^\mathrm{cl}_{\nu} = 10^4\ n_{0\nu}$ (from top to 
bottom in each plot). The colour code is the same as in fig.~1.  }}
\label{fig:cluster}
 \end{center}
 \end{figure*}

\section{Absorption lines in the case of relic neutrino clustering}

 The possible clustering of relic neutrinos onto
dark matter halos has been intensively studied, in particular in the
context of the generation of the UHE cosmic rays through the Z-burst
mechanism~\cite{fargion99,weiler99}. Recent calculations using
Vlasov (Boltzann collisionless) equation have presented revised
estimations of the density profiles and typical spatial extension of
the neutrino clusters~\cite{singh03,ringwald04}. They give
overdensities of the order of $10$-$10^{4}\ n_{0\nu}$ and cluster
scales $ L \sim 0.01 - 1$ Mpc, depending on the neutrino mass, the
mass of the attracting halo and its velocity dispersion (typically ~
200 km/s for a galaxy and ~ 1000 km/s for a galaxy cluster). Limits
to the clustering of neutrinos on large scales are also set by the
Pauli exclusion principle and the limit on the maximum phase-space
density~\cite{Tremaine79}, which imply that only neutrinos with mass
$m_\nu \geq 1$ eV will efficiently cluster on galactic halos ($L_G
\sim 50$ kpc), while neutrinos with $m_\nu \geq 0.1$ eV can cluster
on the much bigger scales associated to halos of (super-)clusters
($L_C \sim 1$ Mpc)~\cite{weiler99,singh03,ringwald04}.

To compute the \UHEnu absorption due to clustered neutrinos, we
substituted $f_\nu(P)$ in eq.(\ref{eq:gamma}) by a modified
Fermi-Dirac distribution,
\begin{equation}
  \label{eq:f-ansatz}
  f_\nu^\mathrm{cl}(P) = \frac{1}{2}\, \frac{e^{-\Phi/T_\nu}+1}{e^{(P-\Phi)/T_\nu} + 1}.
\end{equation}
which parametrises reasonably well the distributions functions
presented in \cite{ringwald04} in function of a single parameter
$\Phi$.  The neutrino density corresponding to eq.\ref{eq:f-ansatz}
is
\begin{equation}
  \label{eq:cluster-n}
  n_\nu^\mathrm{cl} = -\frac{T_\nu^3}{2\pi^2} (1+e^{-\Phi/T_\nu})\,
  \mathrm{Li}_3(-e^{\Phi/T_\nu}),
\end{equation}
where \(\mathrm{Li}_3(x)\) is the trilogarithm function. We also
assume that the cluster density is constant, i.e.
\[\left\{\begin{array}{lcll}
n_\nu &=& N_{cl}\ n_{\nu 0} & r<L_{cl} \\
n_\nu &=& n_{\nu 0} & r>L_{cl}, \\
\end{array}\right.\]
so that $f_\nu(P)$ does not depend on the position. For a given
overdensity factor \(N_\mathrm{cl}\) we then solve $n_\nu^\mathrm{cl} = N_\mathrm{cl} \,n_{\nu 0}$ 
numerically for \(\Phi\).

We then computed the transmission probability for a cluster of relic
neutrinos with mass 0.1 or 1 eV, located between the \UHEnu source
and the observer. As expected, the effect of the thermal motion of
the neutrinos is generally negligible or small due to the relatively
small overdensities achievable and the absence of redshift effect in
$T_\nu$. Fig. \ref{fig:cluster} shows that we have to saturate the
bounds on both parameters $N^{cl}$ and $L_\mathrm{cl}$ to obtain a
significant effect on the depth of the absorption line: for a
maximal overdensity factor $N_\mathrm{cl} = 10^4$, the maximum
absorption probability across the cluster is reduced from $\approx
55 \%$ to $\approx 35 \%$.

\section{Conclusions}
\label{sec:conclusion}
The study of both absorption and emission
features in the spectrum of UHE neutrinos offers probably today's
most promising method for a relatively direct observation of the
cosmic neutrino background. This is especially true in view of the
upcoming generation of detectors, like ICECUBE \cite {icecube},
ANITA \cite{anita}, FORTE \cite{forte}, GLUE \cite{glue}, the Pierre
Auger Observatory \cite{pao},... which are already putting limits on
the \UHEnu flux and starting to constrain the corresponding
theoretical models.

One has however to keep in mind that the thermal motion of the the
relic neutrinos could significantly affect the shape and position of
the absorption dips in the \UHEnu spectrum. From the exploration of
the parameter space currently allowed by astrophysical and
cosmological constraints, we have seen that, even in the regime of
non-relativistic neutrinos, the dips can be broadened and shifted to
lower energies. This will complicate their observation in real
experiments, especially if the \UHEnu source population is
concentrated at small redshifts, producing rather shallow and
extended dips. The shift to lower energies, where neutrino fluxes
are expected to be higher, could in principle increase the detection
potential respect to the case of a relic neutrino at rest with the
same mass. Still, the situation will be more intricate if the
pattern of neutrino mass eigenstates is such that their combined
effect results in a superposition of dips with different depths and
extensions. Considering the small number of events expected in the
\UHEnu detectors at those high energies, it seems rather unrealistic
to resolve the detail of these complex absorption patterns.

As for the case of \UHEnu going through relic neutrino clusters, we
have seen that the effect on the absorption is relatively small
compared to that of \UHEnu that travel on cosmological distances,
since clustering only occurs at small redshifts. For the same
reason, and because only rather heavy neutrinos do cluster, thermal
effects have a limited impact on the shape of the absorption lines
and only result in their attenuation. On the other hand, nearby
clusters of relic neutrinos are expected to play an important role
in the context of the Z-burst mechanism and the emission of UHE
cosmic rays.


\begin{thebibliography}{9}
\bibitem{peebleskolb}
P. J. E. Peebles, {\it Principles of Physical Cosmology}, Princeton
University Press, 1993; E.W. Kolb and M.S. Turner, {\it The Early
Universe}, Addison-Wesley, Redwood City, 1990.
\bibitem{hannestadpastor}
  S.~Hannestad,
  New J.\ Phys.\  {\bf 6} (2004) 108
  [arXiv:hep-ph/0404239]; J.~Lesgourgues and S.~Pastor,
  Phys.\ Rept.\  {\bf 429}, 307 (2006)
  [arXiv:astro-ph/0603494].
\bibitem{eberle}
  B.~Eberle, {\it ``Big bang relic neutrinos and their detection''},
DESY-THESIS-2005-024 
\bibitem{fargion99}D.~Fargion, B.~Mele, and A.~Salis,
Astrophys.\ J.\  {\bf 517}, 725 (1999) [arXiv:astro-ph/9710029].
\bibitem{weiler99}T.~J.~Weiler,
Astropart.\ Phys.\  {\bf 11}, 303 (1999) [arXiv:hep-ph/9710431].
\bibitem{ringwald05} A.~Ringwald, T.~J.~Weiler, and Y.~Y.~Y.~Wong,
  arXiv:astro-ph/0505563.
\bibitem{weiler}
T.~J.~Weiler,
Phys.\ Rev.\ Lett.\  {\bf 49}, 234 (1982) and
Astrophys.\ J.\  {\bf 285}, 495 (1984).
\bibitem{roulet} E.~Roulet,
Phys.\ Rev.\ D {\bf 47}, 5247 (1993).

\bibitem{yoshida}S.~Yoshida,
Astropart.\ Phys.\  {\bf 2} (1994) 187;  S.~Yoshida, H.~y.~Dai,
C.~C.~H.~Jui, and P.~Sommers,
Astrophys.\ J.\  {\bf 479}, 547 (1997) 
\bibitem{ringwald}
B.~Eberle, A.~Ringwald, L.~Song, and T.~J.~Weiler,
Phys.\ Rev.\ D {\bf 70}, 023007 (2004).
\bibitem{quigg}
 G.~Barenboim, O.~Mena
Requejo, and C.~Quigg,
Phys.\ Rev.\ D {\bf 71} (2005) 083002.
\bibitem{pas}
 H.~Pas and T.~J.~Weiler,
  Phys.\ Rev.\ D {\bf 63} (2001) 113015.
  \bibitem{fargion}
  D.~Fargion, P.~G.~De Sanctis Lucentini, M.~Grossi, M.~De 
Santis, and B.~Mele,
  Mem.\ Soc.\ Ast.\ It.\  {\bf 73} (2002) 848.


\bibitem{spergel03}
  D.~N.~Spergel {\it et al.}  [WMAP Collaboration],
  Astrophys.\ J.\ Suppl.\  {\bf 148} (2003) 175
  [arXiv:astro-ph/0302209].
\bibitem{bell} J.~F.~Beacom and N.~F.~Bell,
  Phys.\ Rev.\ D {\bf 65} (2002) 113009.
\bibitem{dolivo} J.~C.~D'Olivo and J.~F.~Nieves,
  Phys.\ Rev.\ D {\bf 52} (1995) 2987.
\bibitem{ourpaper}  J.~C.~D'Olivo, L.~Nellen, S.~Sahu and V.~Van Elewyck,
  Astropart.\ Phys.\  {\bf 25} (2006) 47
  [arXiv:astro-ph/0507333].
  \bibitem{singh03}
S.~Singh and C.~P.~Ma,
Phys.\ Rev.\ D {\bf 67}, 023506 (2003) [arXiv:astro-ph/0208419].
\bibitem{ringwald04}A.~Ringwald and Y.~Y.~Y.~Wong,
JCAP {\bf 0412}, 005 (2004) [arXiv:hep-ph/0408241].
\bibitem{Tremaine79}
  S.~Tremaine and J.~E.~Gunn,
  Phys.\ Rev.\ Lett.\  {\bf 42} (1979) 407.
\bibitem{icecube} J.~Ahrens {\it et al.}  [The IceCube Collaboration],
  Nucl.\ Phys.\ Proc.\ Suppl.\  {\bf 118} (2003) 388.
\bibitem{anita} S.~W.~Barwick {\it et al.}  [ANITA Collaboration],
  Phys.\ Rev.\ Lett.\  {\bf 96} (2006) 171101.
\bibitem{forte}N.~G.~Lehtinen, P.~W.~Gorham, A.~R.~Jacobson and R.~A.~Roussel-Dupre,
  Phys.\ Rev.\ D {\bf 69} (2004) 013008.
\bibitem{glue}P.~W.~Gorham, C.~L.~Hebert, K.~M.~Liewer, C.~J.~Naudet, D.~Saltzberg and D.~Williams,
  Phys.\ Rev.\ Lett.\  {\bf 93} (2004) 041101
\bibitem{pao} X.~Bertou, P.~Billoir, O.~Deligny, C.~Lachaud and A.~Letessier-Selvon,
  Astropart.\ Phys.\  {\bf 17} (2002) 183.
\end{thebibliography}
\end{document}